\documentclass[conference]{IEEEtran}
      
\usepackage{url}
\usepackage{graphicx}  
\usepackage{caption}  
\usepackage{subcaption}    
\usepackage{color}
\usepackage{hyperref}    
\usepackage{mathtools}  
\usepackage{amssymb}  
\usepackage{amsmath}  
\usepackage{amsthm}
\usepackage{algorithm}    
\usepackage{algpseudocode}    
\usepackage{enumerate}    
\usepackage{tabularx}    
\usepackage{multirow}    
\usepackage{setspace}            

\usepackage[keeplastbox]{flushend}
    
\algnewcommand{\LineComment}[1]{\State /*#1*/}
\algrenewcomment[1]{/*#1*/}

\begin{document}

\title{Crossfire Attack Detection using Deep Learning in \\ Software Defined ITS Networks}
\author{
  \IEEEauthorblockN{
    Akash Raj Narayanadoss, Tram Truong-Huu, Purnima Murali Mohan, Mohan Gurusamy
 }
  \IEEEauthorblockA{
    Department of Electrical and Computer Engineering \\
   National University of Singapore, Singapore
  }

\IEEEauthorblockA{
     E-mail: e0254845@u.nus.edu, \{tram.truong-huu, elepumm, elegm\}@nus.edu.sg
 }
}
\maketitle   

\begin{abstract}
Recent developments in intelligent transport systems (ITS) based on smart mobility significantly improves safety and security over roads and highways. ITS networks are comprised of the Internet-connected vehicles (mobile nodes), roadside units (RSU), cellular base stations and conventional core network routers to create a complete data transmission platform that provides real-time traffic information and enable prediction of future traffic conditions. However, the heterogeneity and complexity of the underlying ITS networks raise new challenges in intrusion prevention of mobile network nodes and detection of security attacks due to such highly vulnerable mobile nodes. In this paper, we consider a new type of security attack referred to as crossfire attack, which involves a large number of compromised nodes that generate low intensity traffic in a temporally coordinated fashion such that target links or hosts (victims) are disconnected from the rest of the network. Detection of such attacks is challenging since the attacking traffic flows are indistinguishable from the legitimate flows. With the support of software-defined networking that enables dynamic network monitoring and traffic characteristic extraction, we develop a machine learning model that can learn the temporal correlation among traffic flows traversing in the ITS network, thus differentiating legitimate flows from coordinated attacking flows. We use different deep learning algorithms to train the model and study the performance using Mininet-WiFi emulation platform. The results show that our approach achieves a detection accuracy of at least $80\%$.
\end{abstract}      
 

\section{Introduction}    
\label{sec:intro}  

Vehicles are becoming smarter with the development of communication technologies and  embedded components called on-board units (OBUs) such as cameras, sensors, radars and global positioning devices. The OBUs are connected to the Internet through various radio access technologies such as Dedicated Short Range Communication (DSRC) supported by roadside units (RSUs) and Long Term Evolution (LTE) networks supported by cellular base stations, forming vehicular networks. Vehicular networks  enable a wide range of applications that analyze real-time data captured by OBUs such as road hazards, accidents and traffic density, and then provide guiding instructions to passengers and vehicles for smart mobility. Vehicular networks combined with government road traffic control systems build up the intelligent transport systems (ITS), which significantly improve safety and security of citizens over roads and highways~\cite{meneguette:2018}.  

Despite the advantages of ITS, the presence  of Internet-connected on-board devices (known as Internet of Things or IoT) on the vehicles expose the underlying ITS networks to new security breaches~\cite{lacroix:2014}. IoT devices have low computing capacity, making them difficult or even impossible to deploy sophisticated security mechanisms. A large number of highly vulnerable on-board IoT devices could be compromised to launch large-scale attacks such as Distributed Denial of Service (DDoS) where bots are installed in the compromised devices that simultaneously send traffic to a targeted victim. Also due to the increased heterogeneity and complexity of ITS networks, new challenges arise in design and implementation of intrusion and attack detection mechanisms.

In this paper, we consider a powerful attack that is launched by coordinating a large number of compromised devices, called crossfire attacks~\cite{kang:2013}. Crossfire attacks aim at isolating a network region from the remaining partition of the network through coordinated flooding of the \textit{pivotal links}. Pivotal links are the links that connect the target region (e.g., a city) to the rest of the world. To remain undetected, the attacker chooses a set of \textit{decoy servers} surrounding the target region that are the destination of the attacking traffic originating from the bots. To launch crossfire attacks,  bots installed in multiple compromised devices are programmed to temporally coordinate and alternatively send low intensity traffic flows to the decoy servers through the pivotal links such that the pivotal links are congested during an \textit{attacking window} (e.g., $30$ minutes). This attacking window can also be extended indefinitely by frequently changing the bots, pivotal links and decoy servers for the target region. A victim server located in the target region is therefore unreachable and becomes unavailable. These low intensity traffic flows are indistinguishable from the legitimate flows since they come from valid IP address and have similar traffic characteristics as the legitimate ones. However, the cumulative amount of traffic generated by the bots in the attacking window causes bottleneck problem on the pivotal link(s). Several links could be flooded, leading to disconnection among network partitions if the links are pivotal. Detection of such crossfire attacks in ITS is even more challenging since existing approaches for DoS/DDoS attack detection are no longer applicable~\cite{liaskos:2016,xue:2018}.    

The emergence of the software-defined networking (SDN) paradigm enables network programmability and monitoring with a separate controller that can run sophisticated algorithms for traffic engineering and security purposes. Application of SDN to ITS networks has shown improved performance in network resource management~\cite{Fontes:2017}. In this paper, we leverage on the SDN capability to develop a machine learning approach for detection of crossfire attacks in ITS. The SDN controller collects traffic measurements of flows destined to different hosts in the network. The machine learning model deployed in the controller helps in learning the temporal correlation among the traffic flows, thus being able to differentiate between malicious and legitimate flows. We use deep learning techniques based on artificial neural networks to train the learning model since deep learning has better ability to characterize the inherent relationships between the inputs and outputs of networks without human involvement. It also guarantees a finite convergence for linearly-separable data. We evaluate the performance of the proposed approach using Mininet-WiFi emulation platform that allows us to emulate the mobility of vehicles with on-board IoT devices.  
  
The rest of the paper is organized as follows. We review the related
work in Section~\ref{sec:related_work}. We present our proposed
approach in Section~\ref{sec:approach}. In Section~\ref{sec:deep_learning}, we present different deep learning algorithms used for training the learning model. We present performance evaluation in Section~\ref{sec:results} before we conclude the paper
in Section~\ref{sec:conclusion}.

\section{Related Work}
\label{sec:related_work}

There has been research work carried out in the field of crossfire attacks~\cite{kang:2013,studer:2009}. In~\cite{studer:2009}, the authors introduced the Coremelt attack, in which bots generate traffic among them to cause congestion on the target links. Crossfire attacks have been presented in~\cite{kang:2013} wherein the bots coordinate and generate traffic towards decoy servers in the network.

A number of works have been carried out in the literature for crossfire attack detection and mitigation~\cite{gkounis:2014,Xue:2014,shende:2015}. In~\cite{gkounis:2014}, the authors presented an approach to defeat crossfire attacks using a centralized flow-level control and monitoring. In this model, the defender tries to keep the network running without any congestion by performing load balancing among links and paths. The defender also records the sources that cause traffic congestion and rate-limits the traffic if the cumulative traffic threshold is exceeded. Using a similar approach, in~\cite{shende:2015}, the authors proposed to divide the network into multiple domains and perform traffic rerouting at the local-domain level or inter-domain level. However, these works do not consider the temporal correlation between traffic flows and the use of threshold approach could affect legitimate flows if their traffic volume also exceeds the threshold. 
Our work differs from the above works in that we aim at detecting crossfire attacks based on traffic behavior and the  temporal correlation between them. In~\cite{Xue:2014}, the authors developed LinkScope, a system used to detect link flooding attacks and to locate the target area or the link. The proposed system in~\cite{Xue:2014} used hop-by-hop and end-to-end network measurement techniques that probe the network with testing packets and measure the traffic characteristics to detect the performance degradation. This adds redundant traffic to the network. Whereas, we use machine learning approaches that capture the traffic behavior to detect the attacks.

Machine learning algorithms have also been used in other works for DDoS detection and mitigation leveraging on the features of SDN. In~\cite{Braga:2010}, the authors presented their approach that uses Self Organizing Maps (SOMs) to analyze traffic characteristics captured by NOX controller in an OpenFlow-enabled network.  In~\cite{Ashraf:2014}, the authors developed machine learning approaches for mitigating DDoS attacks in software-defined networks. The SDN controller performs traffic analysis and defines mitigation rules that are installed in the switches. In our work, we propose to use machine learning approach to detect crossfire attacks in the context of ITS.

  
  
\section{ITS Networks and Proposed Approach}
\label{sec:approach}

In this section, we present an overview on software-defined intelligent transport system networks and our proposed approach for coordinated attack detection.

\subsection{Software-defined ITS Networks}

 \begin{figure}
 \centering
   \includegraphics[width=0.49\textwidth]{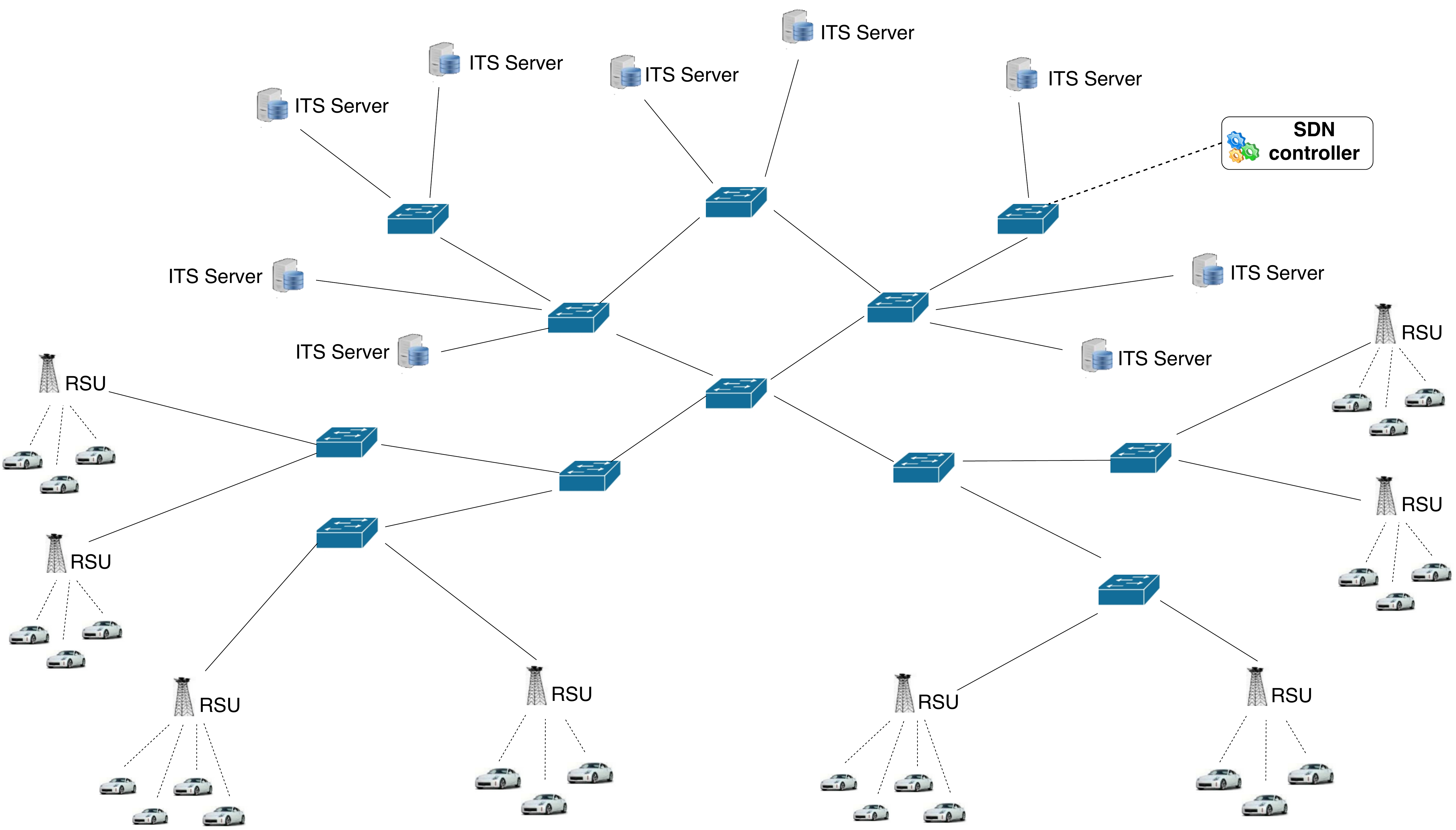}
   \caption{An intelligent transport system network.}
   \label{fig:net_architecture}
 \vspace{-1.5ex}
 \end{figure}

Fig.~\ref{fig:net_architecture} presents a simplified architecture of a software-defined ITS network. All the vehicles moving on the road communicate with RSUs through wireless communication protocols to exchange information with other vehicles or with the system. The RSUs are used as relay nodes or access points for the vehicles to forward traffic to other parts of the network. They are connected to the core network and ITS servers with routers and switches through wired network. With the support of SDN, we assume that all the RSUs and switches in the network are OpenFlow-enabled such that an SDN controller is required to manage the network: installing rules in the network nodes for traffic engineering and monitoring. 

Apart from the vehicle-RSU communications, there can be hop-by-hop mode of communication among the vehicles (V2V communication). This allows the vehicles to exchange information without relaying to RSUs and core network. However, due to different velocity of vehicles, vehicle density and limited spectrum allocation, this hop-by-hop communication mode has poor performance and it is not reliable as compared to relaying to RSUs. Vehicles can also communicate through cellular networks with the support of base stations connected to the core network. In any case, the traffic destined to the servers (where ITS services are located) will be captured by the SDN controller at every OpenFlow-enabled switch.   

\subsection{Proposed Approach}

Crossfire attack detection can be done at different locations: at the source of attack (bots), at the destination (decoy servers) or at the pivotal links. While detection at the source of the attack may be impossible due to the spatial distribution of bots in the network, detection at the decoy servers is also difficult since many decoy servers could be there in the network. Furthermore, the decoy servers may not be in the targeted area of attack, leading to high network resource consumption of attacking traffic. 

Leveraging on the SDN capability, we propose to detect the attack at the pivotal link level. The SDN controller probes traffic characteristics from every switch port (link-level monitoring). We develop the deep learning techniques that will be run on top of the SDN controller to analyze the traffic characteristics captured from the network links. The deep learning techniques based on artificial neural networks can learn traffic behavior, determine temporal and spatial correlations among traffic flows in the network, thus being able to detect whether the network is under an attack or not. Three traffic characteristics are selected as features in our model:
\begin{itemize}
\item \textit{Number of flows} represents the vehicle density and possibly the number of bots installed in the vehicles. 

\item \textit{Aggregate flow size} represents the traffic volume sent through a link. We note that a sudden change (increase) in the flow size can be caught by firewalls or rule-based detection approaches, but bots in crossfire attacks gradually increase flow size to flood the pivotal links in the attacking window. 
\item \textit{Timestamp} is an important feature that represents the temporal correlation among flows. If bots alternatively generate traffic and send to decoy servers, total traffic traversing a link could create congestion even though the individual flow size is low.  
\end{itemize}

To realize the proposed approach, two practical implementations can be considered. The first possible implementation is that the controller periodically captures traffic measurements and performs analysis. The higher the frequency of extraction of traffic measurements, the faster the attack detection. The second possible implementation is to use an event-driven approach to trigger link congestion. When a switch experiences abnormal behavior of network traffic such as high packet loss, it sends a request to the controller along with traffic measurements for attack detection. In the next section, we present deep learning techniques that are able to detect crossfire attacks based on traffic measurements discussed above.
  
\section{Deep Learning Algorithms for Learning Model Training}
\label{sec:deep_learning}
  
In this section, we present several deep learning algorithms used to train the learning model that detects the correlation among traffic flows sent from vehicles. Deep learning with multi-layer neural networks has the ability of feature learning in a short time of training. In this paper, we use Artificial Neural Networks (ANN), Convolutional Neural Networks (CNN) and Long Short-Term Memory (LSTM) networks.

\subsection{Artificial Neural Networks}

An artificial neural network depicts the neural distribution of an animal brain. ANN is a collection of units called artificial neurons. The edges connect the artificial neurons. The learning procedure can be adjusted with the help of the weights of the edges. Each connection acts like a synapses in brain that transmits signal from one artificial neuron to another. Depending on the signal strength, the weight of the neuron may get reduced or increased. The artificial neurons are aggregated into layers. Apart from the neurons at the input layer that is fed by the data points there may be also multiple hidden layers of artificial neurons that perform different kinds of transformations on their inputs.

\begin{figure}[t]    
\centering
  \includegraphics[width=0.48\textwidth]{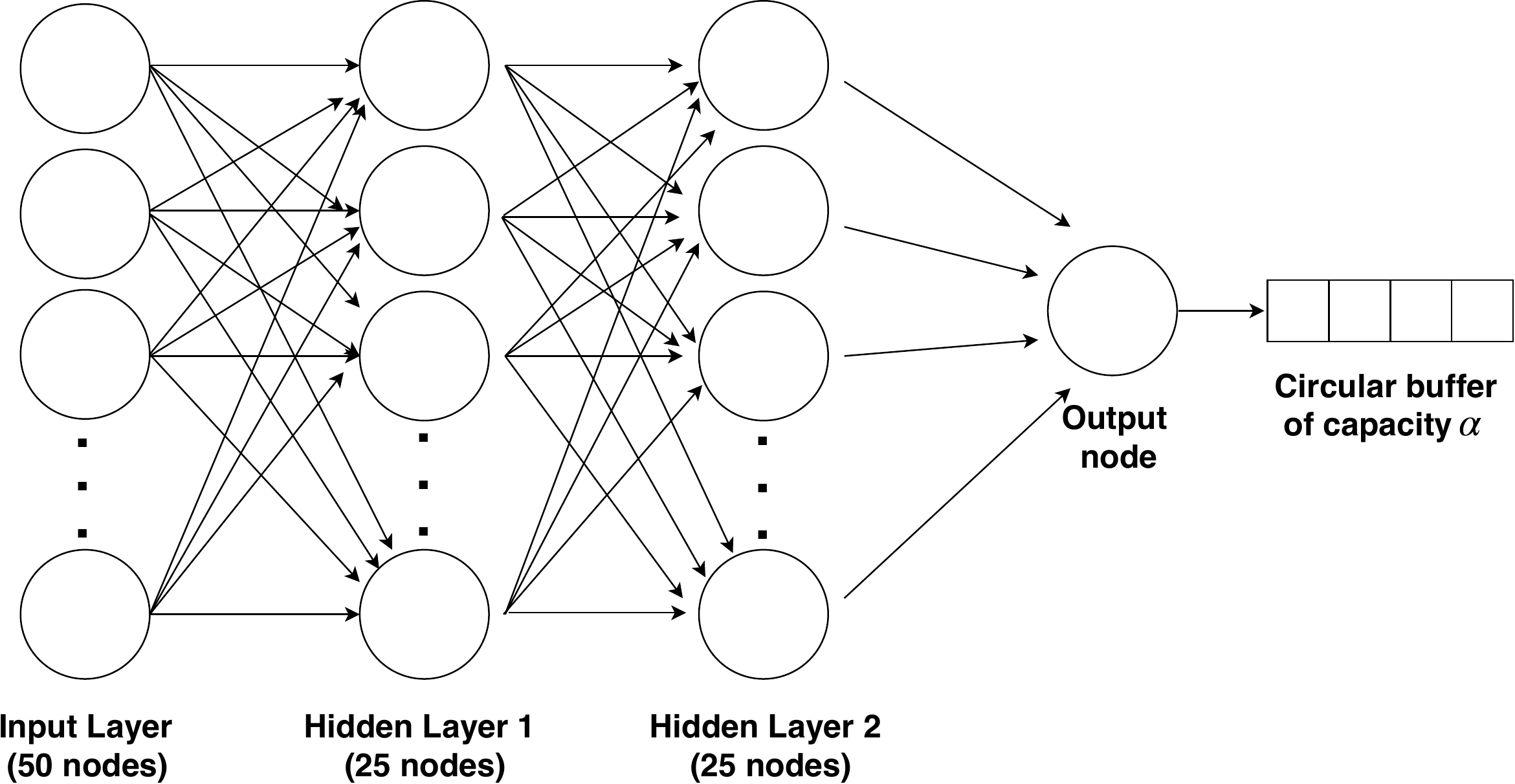}
  \caption{Artificial neural network used in our work.}
  \label{fig:ann}
  \vspace{-1.5ex}    
\end{figure}

Fig.~\ref{fig:ann} depicts an ANN used in our work. The input layer consists of $50$ nodes. The first node takes the timestamp of data points. Each of remaining node pairs takes the traffic features (number of flows and aggregate flow size) of a link monitored. With this design, our model can monitor and detect the attack on $24$ pivotal links simultaneously. It is worth mentioning that a larger ANN will be required if the protected network has large number of links. In order to increase the efficiency we use two hidden layers, each containing $25$ nodes. The activation function used for the hidden layers is the Rectifier function~\cite{pmlr-v15-glorot11a}. The output layer has only one node that indicates whether or not a network is under an attack. The activation function used for the output node is the sigmoid function~\cite{Haykin:1999}. We use Adam optimizer as the gradient descent to optimize network weights~\cite{Kingma:2015}. We use a circular buffer of size $\alpha$ that indicates the number of times the traffic is consecutively detected as attacking traffic. If the buffer is full, the network is detected to be under an attacking window. 

\subsection{Convolutional Neural Networks}

\begin{figure}[t]  
\centering
  \includegraphics[width=0.48\textwidth]{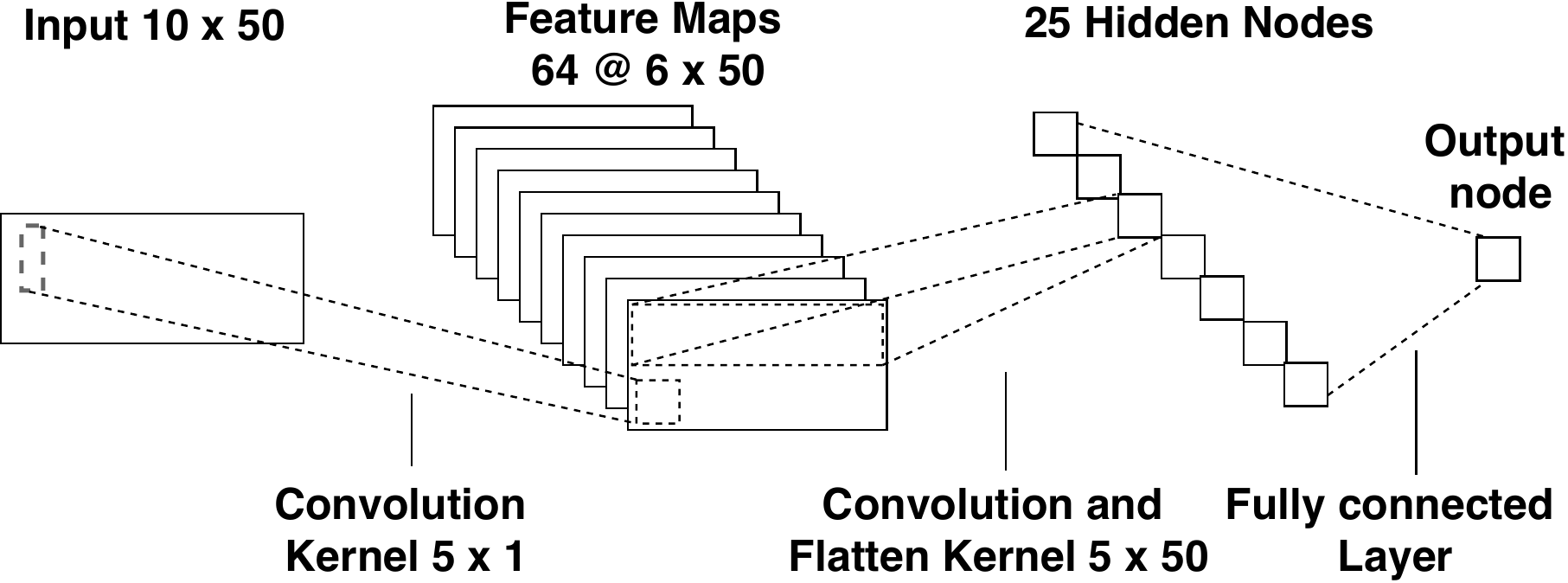}
  \caption{Convolutional neural network used in our work.}
  \label{fig:cnn}
    \vspace{-1ex}
  \end{figure}

A convolutional neural network (depicted in Fig.~\ref{fig:cnn}) is a variation of multi-layer perceptrons that have been designed to reduce the efforts of data pre-processing. Each convolutional neuron processes the data that is within its receptive field. In our work, the traffic measurements (dataset) captured from the network is organized as a $2$-dimensional array. Each row represents the traffic measurement at a particular timestamp. A pair of columns contain the number of flows and aggregate flow size of a link in the network. We consider a time window with $10$ measurements at $10$ different timestamps. Similar to ANN, we define threshold $\alpha$ such that the network is detected to be under an attack if the attacking traffic is detected more than $\alpha$ times. The CNN is structured with two separate convolutional steps. The first operation is a convolutional filter that spans only the time axis, e.g., it processes one row of the dataset at a time. This learning step referred to as temporal filter allows the defender to learn the attacking traffic pattern over time. The second learning step referred to as spatial filter spans across all the monitored links for multiple timestamps, i.e., considering a few rows in the $2$-dimensional array. This learning step aims at detecting the correlation between different links as they forward the attack traffic towards the victim. We also use a Rectifier function for the activation function in CNN. A fully-connected layer is formed with Adam optimizer that results in a binary value which determines whether or not the network is under attack.

\subsection{Long Short-Term Memory Network}              

\begin{figure}[t]                
\centering
  \includegraphics[width=0.48\textwidth]{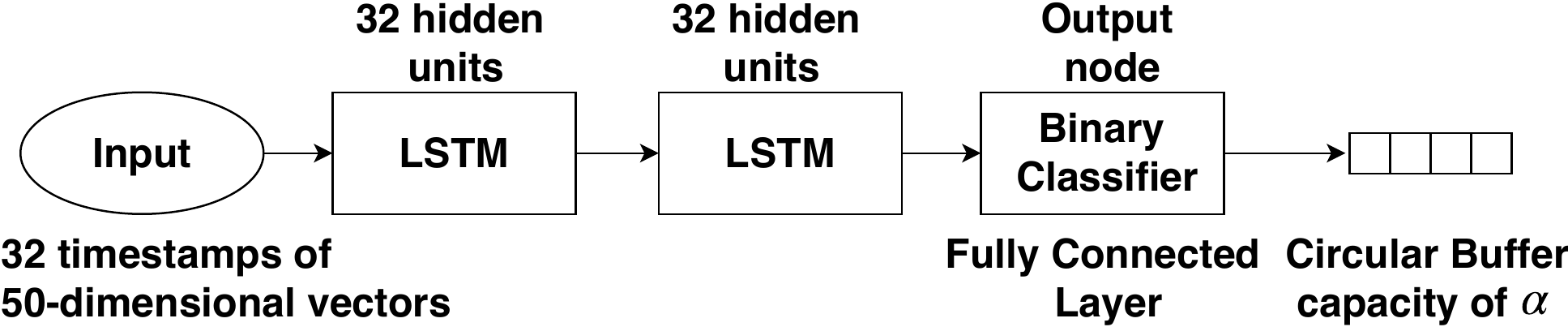}
  \caption{LSTM network used in our work.}
  \label{fig:lstm}
    \vspace{-2.5ex}
\end{figure}

Long Short-Term Memory Network~\cite{Sepp:1997} has been developed to overcome the drawback of recurrent neural network on the long term dependency problems. It is used as  a sequence classifier for detecting whether the traffic at each timestamp is attacking traffic or not. As depicted in Fig.~\ref{fig:lstm}, the LSTM network has two consecutive LSTM cells which forms a stacked LSTM configuration. The inputs to the LSTM network are windows of $32$ timestamps of $50$-dimensional vectors, which contain number of flows and aggregate flow size of $25$ links in the network. Each LSTM cell has $32$ hidden units. With the help of the binary classifier, the output sample from the stacked LSTM is classified into an attack or non-attack class by a fully-connected layer. Adam optimizer is used to train the network with a learning rate $0.001$ till the early stopping condition. We also use a circular buffer of size $\alpha$ that indicates the number of times the traffic is consecutively detected as attacking traffic. If the buffer is full, the network is detected to be under an attacking window. 

\section{Performance Study}    
\label{sec:results}

\subsection{Experimental Setting}

We implement the proposed approach and carry out experiments on Mininet-WiFi platform to evaluate its performance. Mininet-WiFi is an OpenFlow-enabled network emulator based on Linux LXC containers~\cite{Fontes:2017}. It allows us to create an ITS network as depicted in Fig.~\ref{fig:net_architecture} with mobile nodes representing vehicles that follow a specific trajectory and speed. To generate a Crossfire attack, we direct low intensity flows from the vehicles to the hosts. We use \textit{iperf3} to generate traffic in both normal working scenarios and attacking scenarios. In normal working scenarios, the vehicles generate a background traffic with a traffic rate in the range $[600, 1700]$ Kbps. During the attacking window, additional traffic is added along with the background traffic. The compromised vehicles gradually increase the intensity of the traffic in the range $[40, 300]$ Kbps and alternatively send the traffic to decoy servers. We collect $7000$ data-points both in normal and attack scenarios. The total emulation time is $1$ hour. 

We use the following performance metrics to evaluate the performance
of different machine learning algorithms:
  
\begin{itemize}
\item Precision: Ratio of the number of data-points correctly classified as attacking traffic over the total number of data-points predicted as attacking traffic. The precision value is computed as
  follows: 
\begin{equation}
	\mathcal{P} = \dfrac{T_P}{T_P +F_P} 
\label{eq:precision}
\end{equation}
where $\mathcal{P}$ is the precision value, $T_P$ is the number of
``true positives''	and $F_P$ is the number of ``false
positives''. 

\item Recall:  Ratio of the number of data-points correctly classified as attacking traffic over the total number of data-points of actual attacking traffic flows. The
  recall value is given by: 
\begin{equation}
	\mathcal{R} = \dfrac{T_P}{T_P +F_N} 
\label{eq:recall}
\end{equation}
where $F_N$ is the number of ``false negatives''.	

\item $F_1$-Score: The $F_1$-Score is the harmonic average of the
  precision and recall values. It takes a value in the range $[0,1]$. Higher the value of $F_1$-Score, better the performance of  the machine learning technique, i.e., we obtain perfect precision and recall values when $F_1$-Score reaches $1$. It is computed as follows:  
\begin{equation}
	F_{1}\text{-Score} = \dfrac{2\mathcal{P}\mathcal{R}}{\mathcal{P} +\mathcal{R}}. 
\label{eq:f1-score}
\end{equation}
\end{itemize}

\begin{figure*}[t]      
\centering	
\begin{subfigure}[b]{0.242\textwidth}  
\centering
\includegraphics[width=\textwidth]{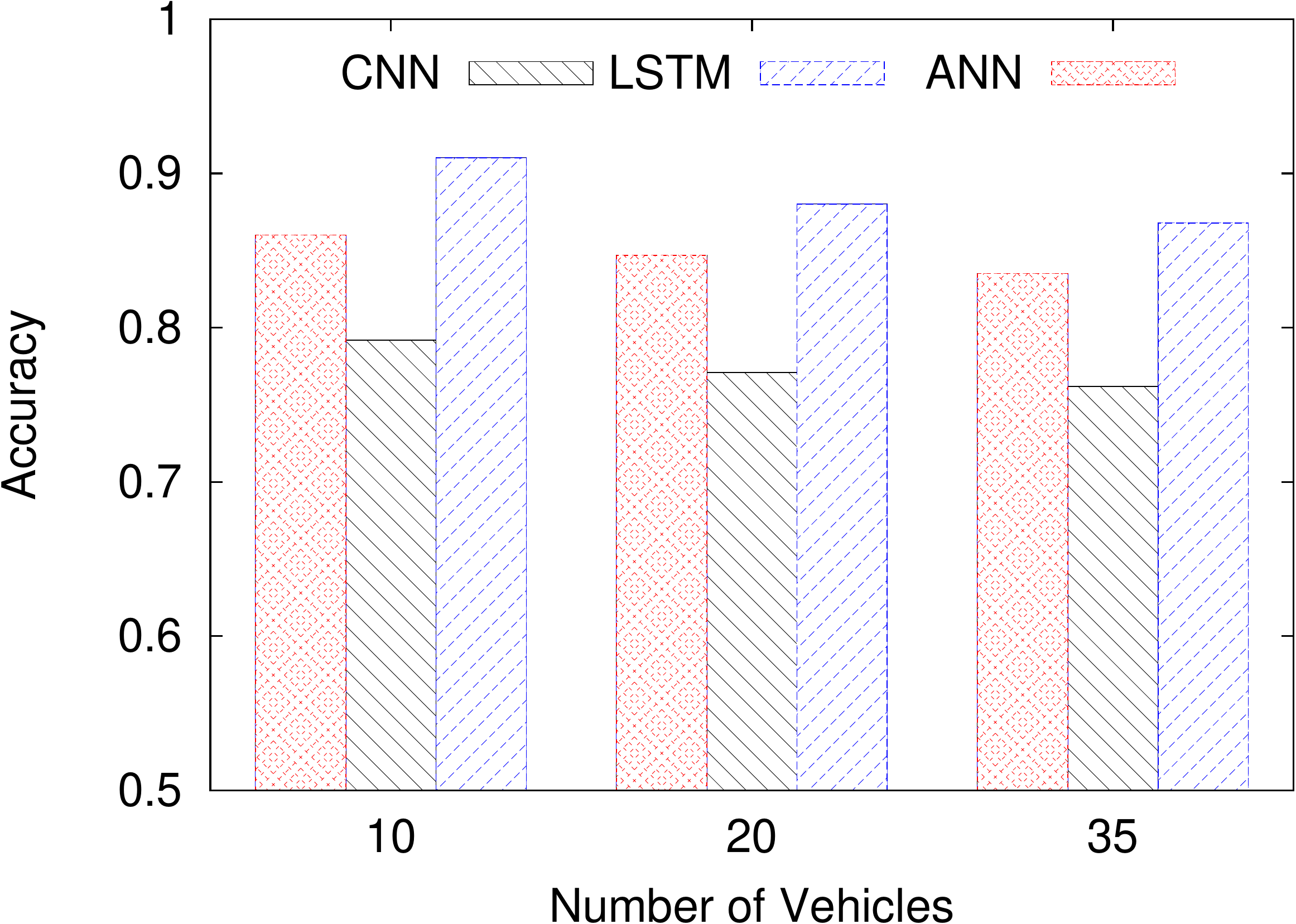}
  \caption{Accuracy.}
  \label{fig:acc_netsize}
\end{subfigure}
\begin{subfigure}[b]{0.242\textwidth}
\centering
 \includegraphics[width=\textwidth]{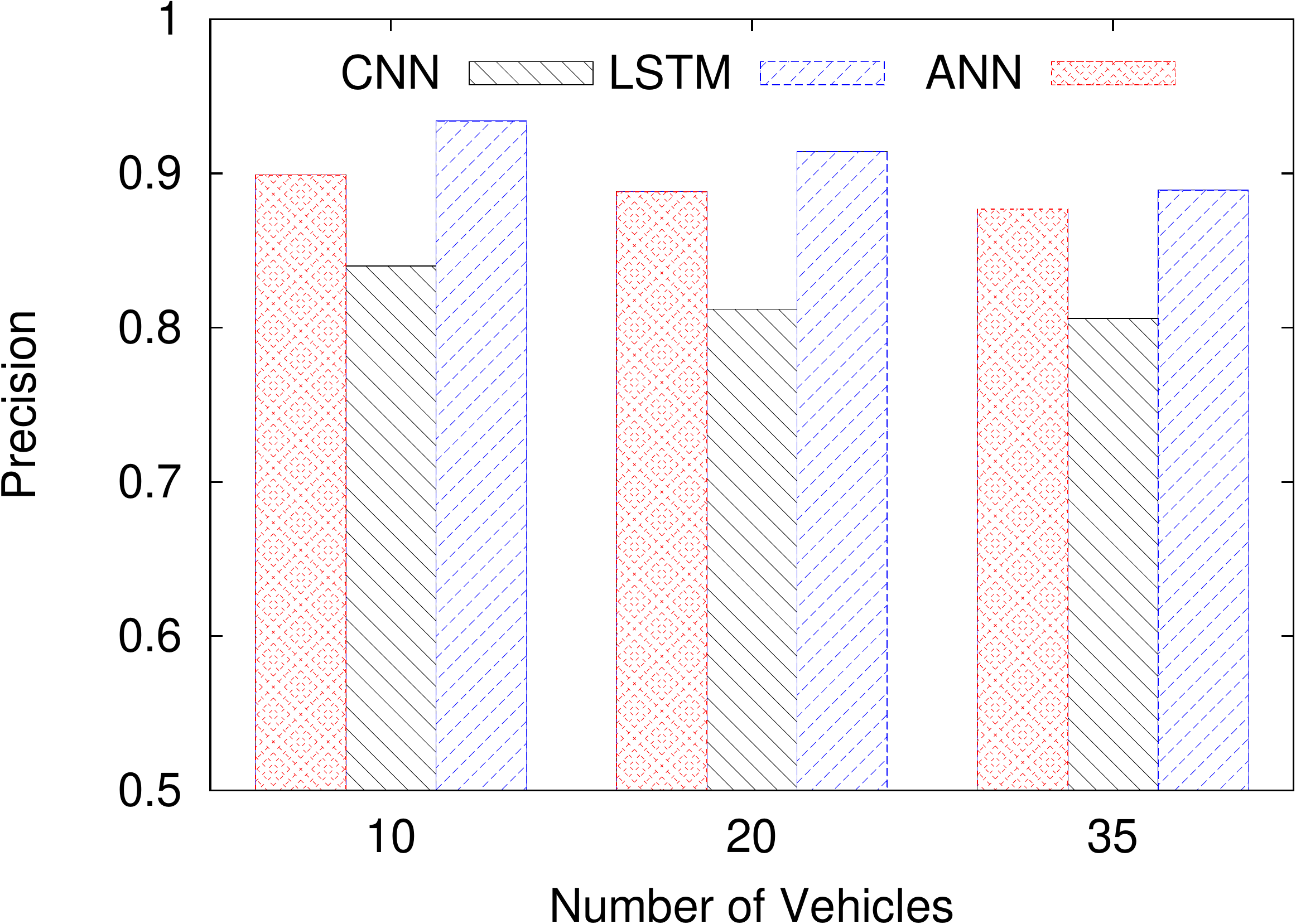}
  \caption{Precision.}
\label{fig:precision_netsize}
\end{subfigure}  
\begin{subfigure}[b]{0.242\textwidth}      
\centering
\includegraphics[width=\textwidth]{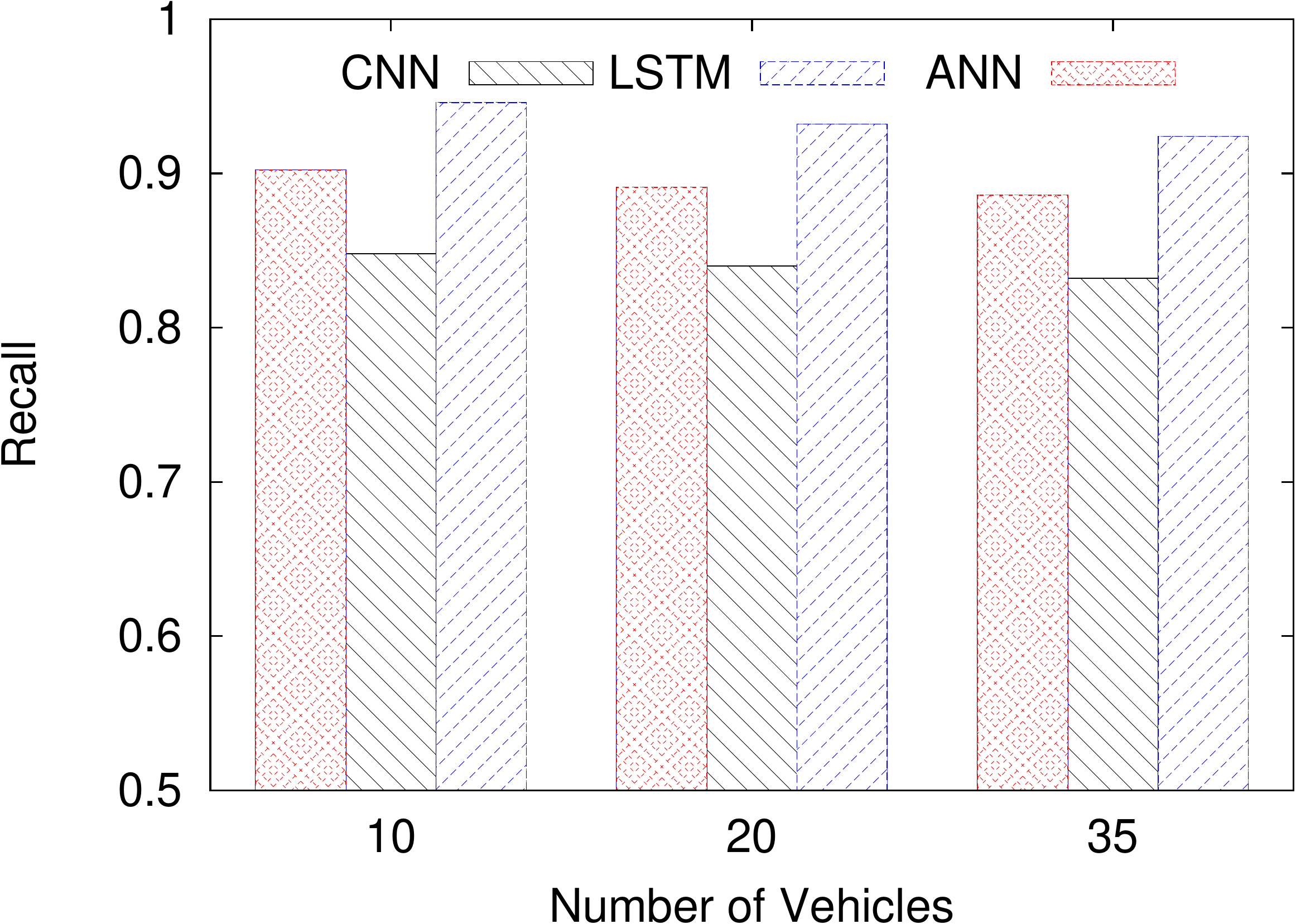}
  \caption{Recall.}
\label{fig:recall_netsize}
\end{subfigure}
\begin{subfigure}[b]{0.242\textwidth}      
\centering
\includegraphics[width=\textwidth]{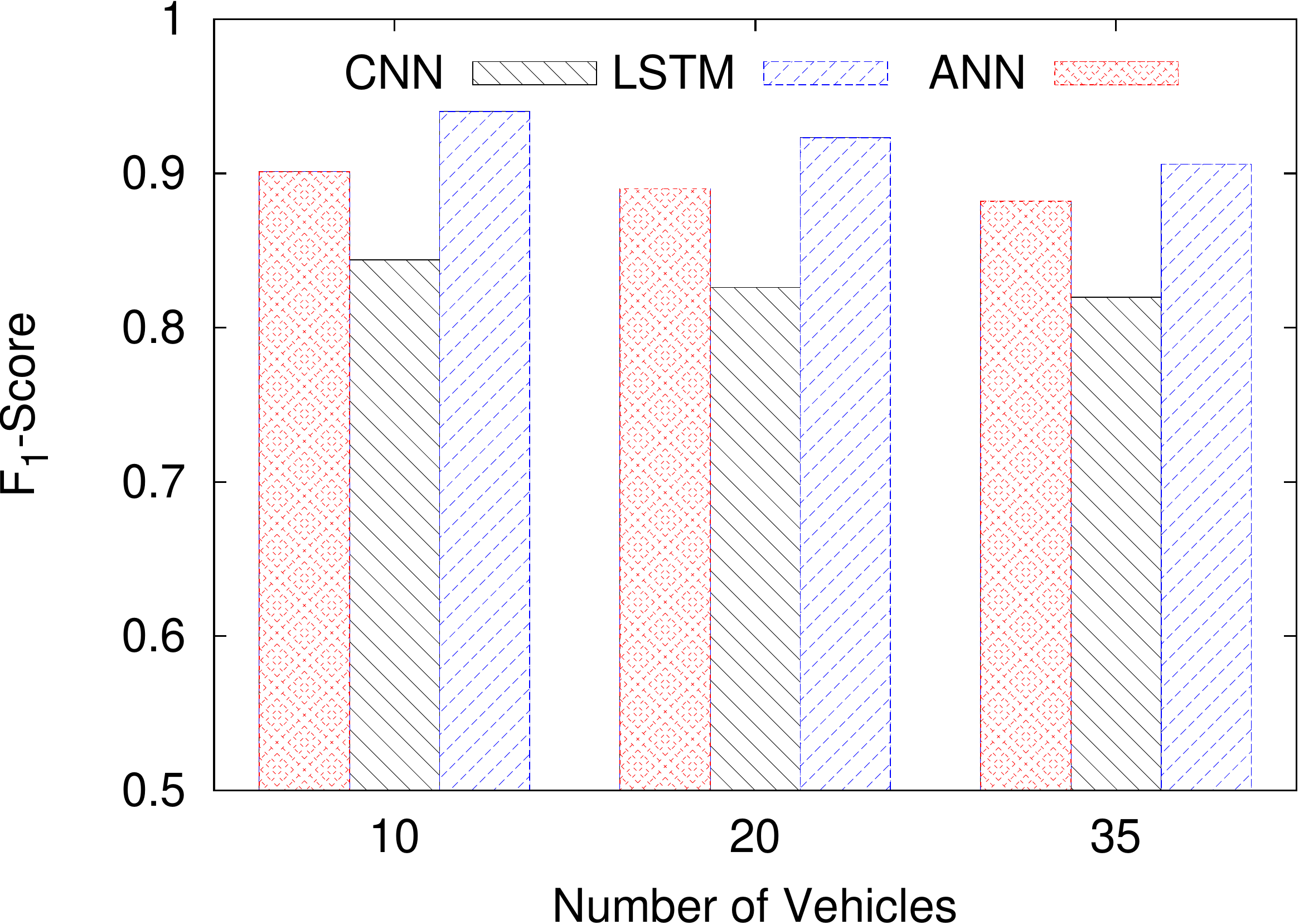}
  \caption{$F_1$-Score.}
\label{fig:f1score_netsize}
\end{subfigure}
\caption{Performance of proposed approach for different number vehicles.}   
\label{fig:performance_netsize}  
\vspace{-1ex}
\end{figure*}

\begin{figure*}[t]      
\centering	
\begin{subfigure}[b]{0.242\textwidth}  
\centering
\includegraphics[width=\textwidth]{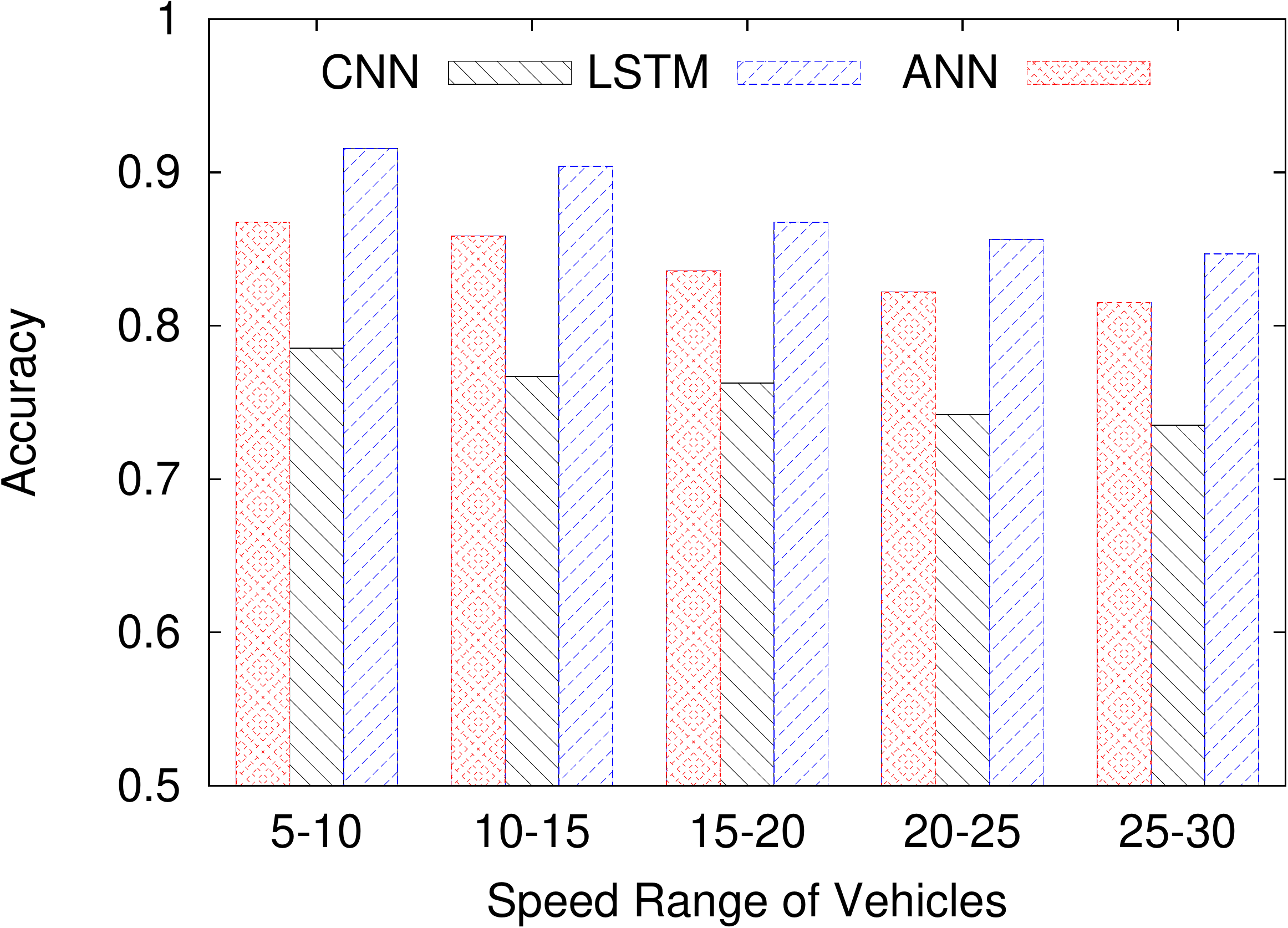}
  \caption{Accuracy.}
  \label{fig:acc_speed}
\end{subfigure}
\begin{subfigure}[b]{0.242\textwidth}
\centering
 \includegraphics[width=\textwidth]{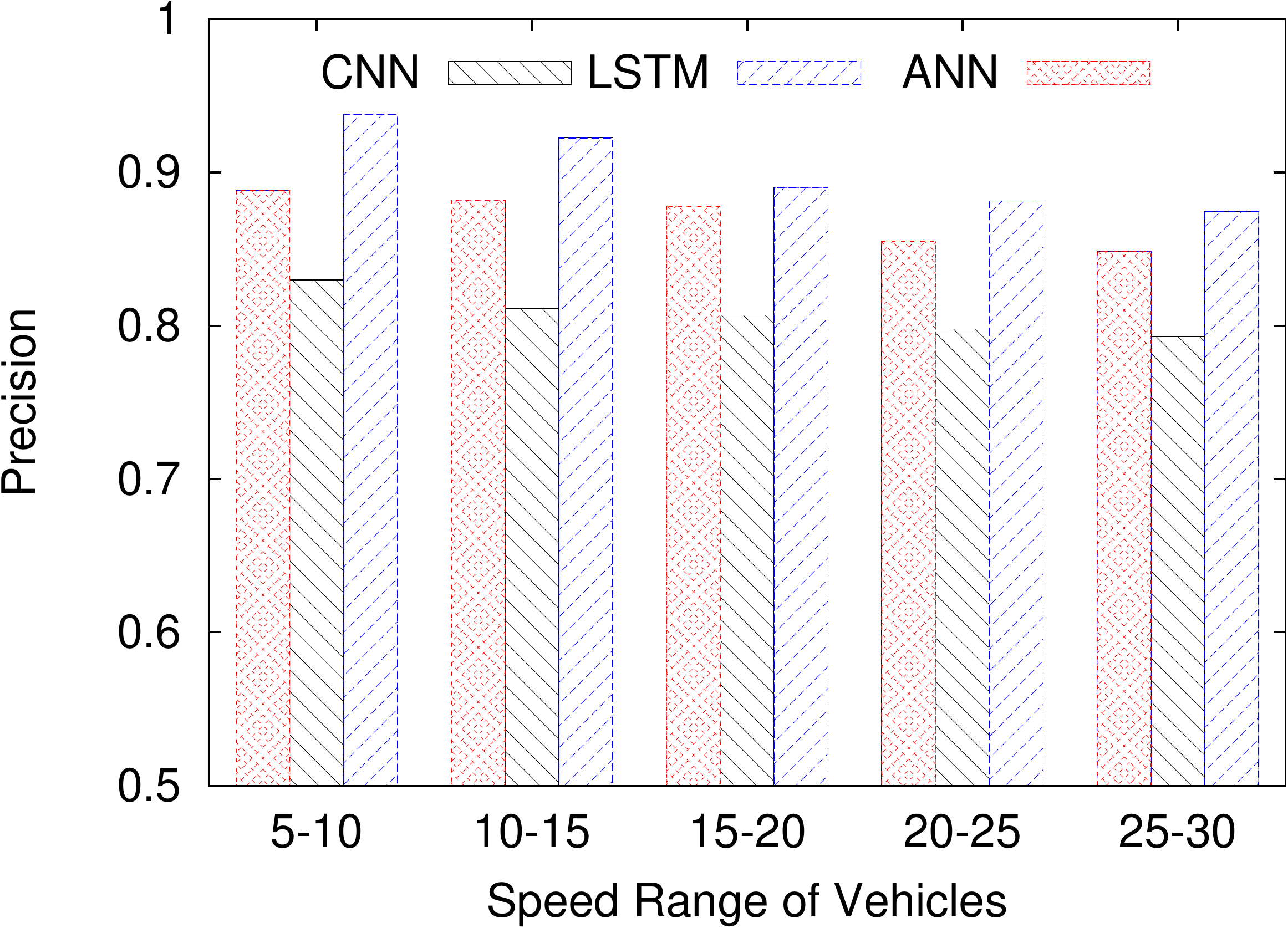}
  \caption{Precision.}
\label{fig:precision_speed}
\end{subfigure}  
\begin{subfigure}[b]{0.242\textwidth}      
\centering
\includegraphics[width=\textwidth]{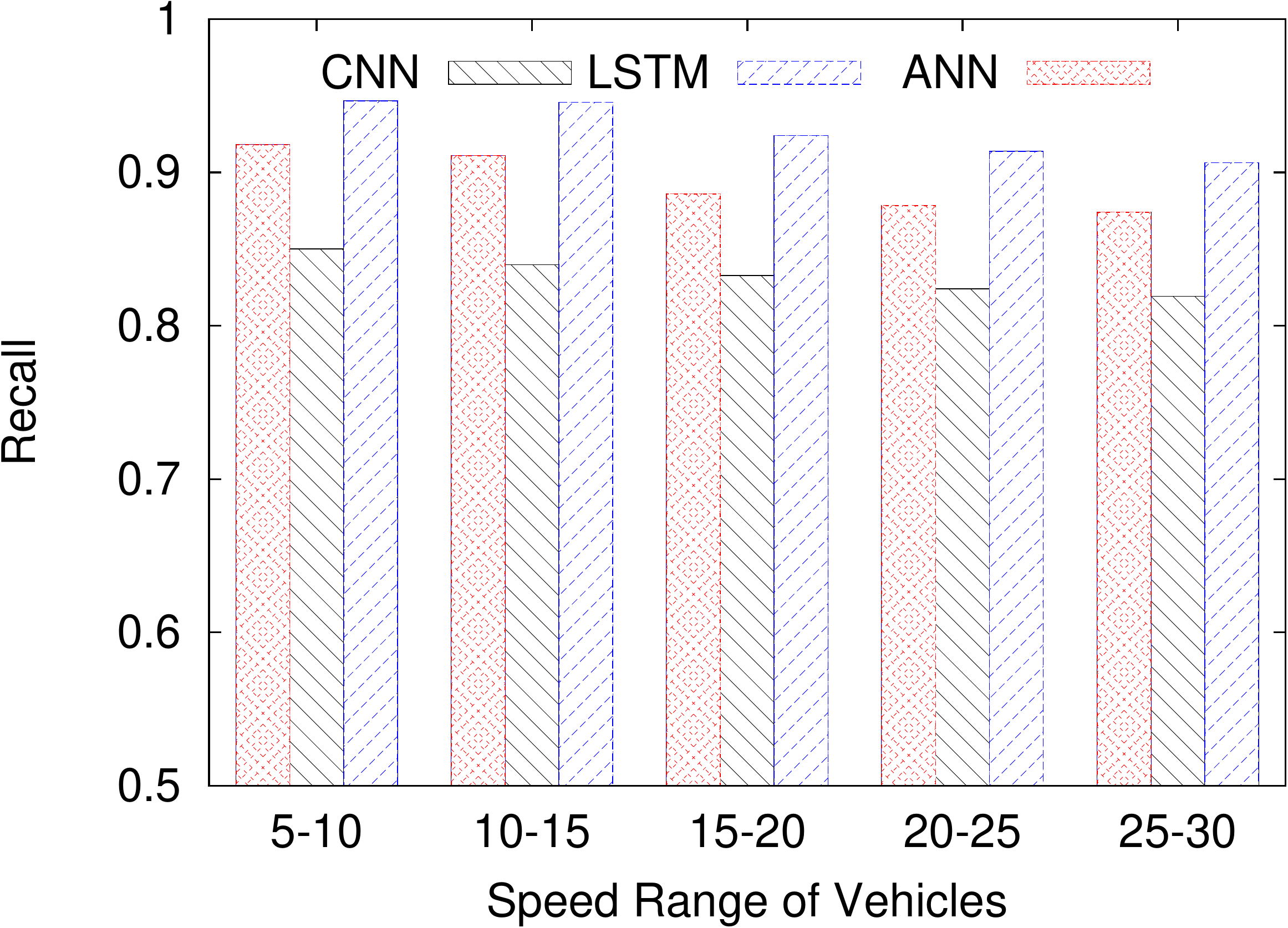}
  \caption{Recall.}
\label{fig:recall_speed}
\end{subfigure}
\begin{subfigure}[b]{0.242\textwidth}      
\centering  
\includegraphics[width=\textwidth]{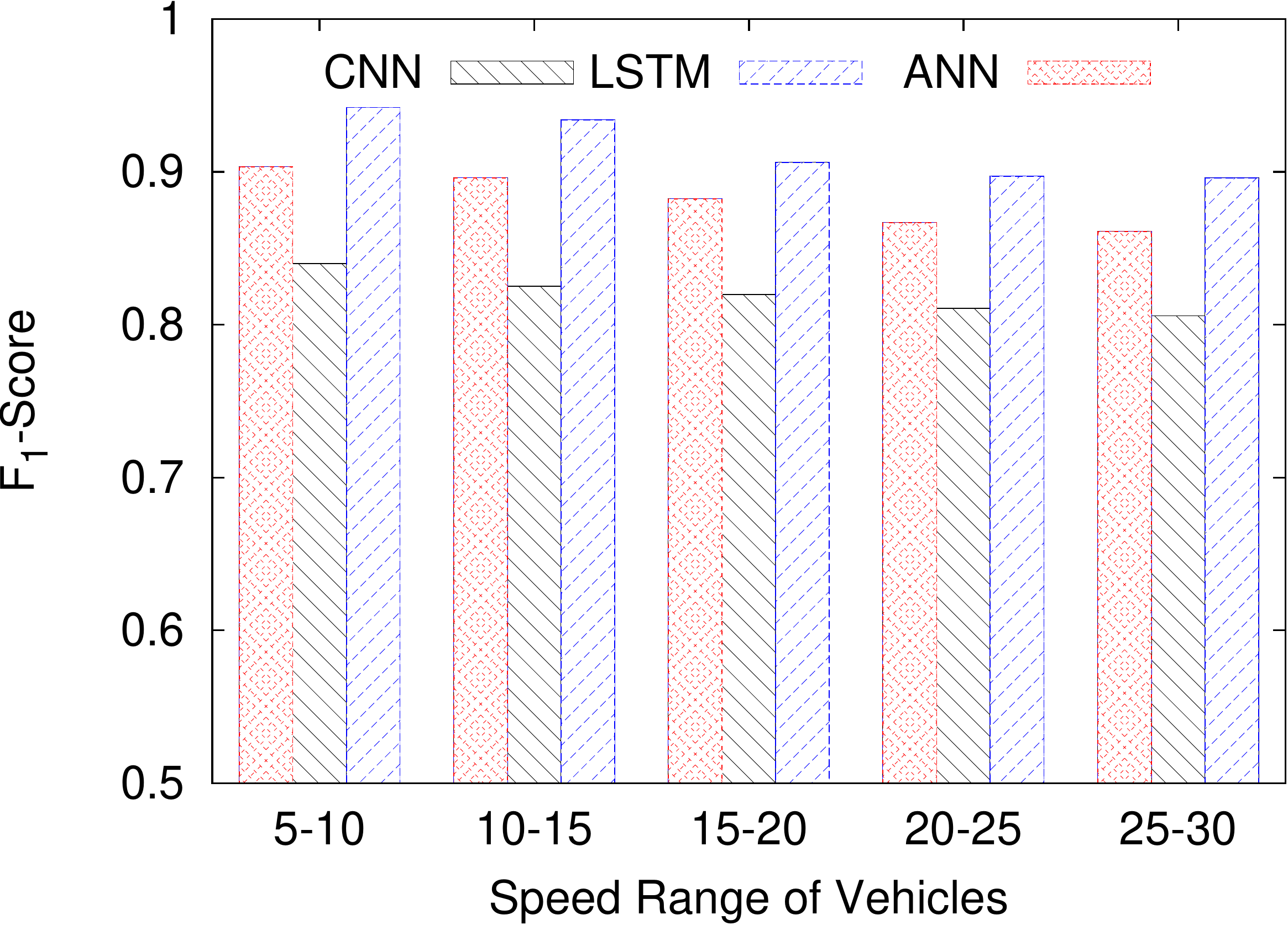}
  \caption{$F_1$-Score.}
\label{fig:f1score_speed}
\end{subfigure}
\caption{Performance of proposed approach for different speed ranges of vehicles.}   
\label{fig:performance_speeds}  
\vspace{-2ex}
\end{figure*} 
\subsection{Analysis of Results}

\subsubsection{Impact of Number of Vehicles}

In the first experiment, we evaluate the impact of the number of vehicles on the performance of the proposed approach. In Fig.~\ref{fig:performance_netsize}, we present all four performance metrics for different network size, i.e., the number of vehicles present in the network. The results show that the performance of the proposed approach degrades with the increasing number of vehicles. This is because large number of  vehicles (implicitly meaning large number of traffic flows) makes it difficult to detect the correct temporal and spatial correlations among flows. Many legitimate flows could be detected as part of attacking traffic as they may have temporal correlation with other attacking flows. In any cases, the results show that using LSTM achieves the best performance with $87\%$ of accuracy when there are $30$ vehicles in the network and up to $91\%$ of accuracy when there are $10$ vehicles in the network. This is because LSTM has better capability of bridging long time lags in time series that contains traffic characteristics of the network. Even though CNN has the lowest performance among the deep learning techniques, it achieves at least $76\%$ with the densest network. 

\subsubsection{Impact of Vehicle Speed}

In the second experiment, we evaluate the impact of vehicle speed on the performance of the proposed approach. We define different speed ranges (in m/s) with each vehicle randomly choosing a speed in that range. In Fig.~\ref{fig:performance_speeds}, we present the performance of the proposed approach. The experimental results show that the performance of the proposed approach slightly degrades when vehicles increase their speed. Due to the wireless connection between vehicles and RSUs, the faster the speed of vehicles, the more the failures in connection establishment between them. These connection failures lead to the communication delay and disorder of arrivals and timestamps of packets at the RSUs where traffic features are extracted. Temporal correlation among packets is therefore much more difficult to determine and achieve high performance. We note that LSTM with its ability of time series analysis, it always achieves the highest performance with $85\%$ of accuracy when vehicles move with the highest speed. It is worth mentioning that LSTM achieves at least $90\%$ in terms of other performance metrics.

\subsubsection{Impact of Attacking Traffic Threshold}

\begin{figure}[t]  
\centering
  \includegraphics[width=0.35\textwidth]{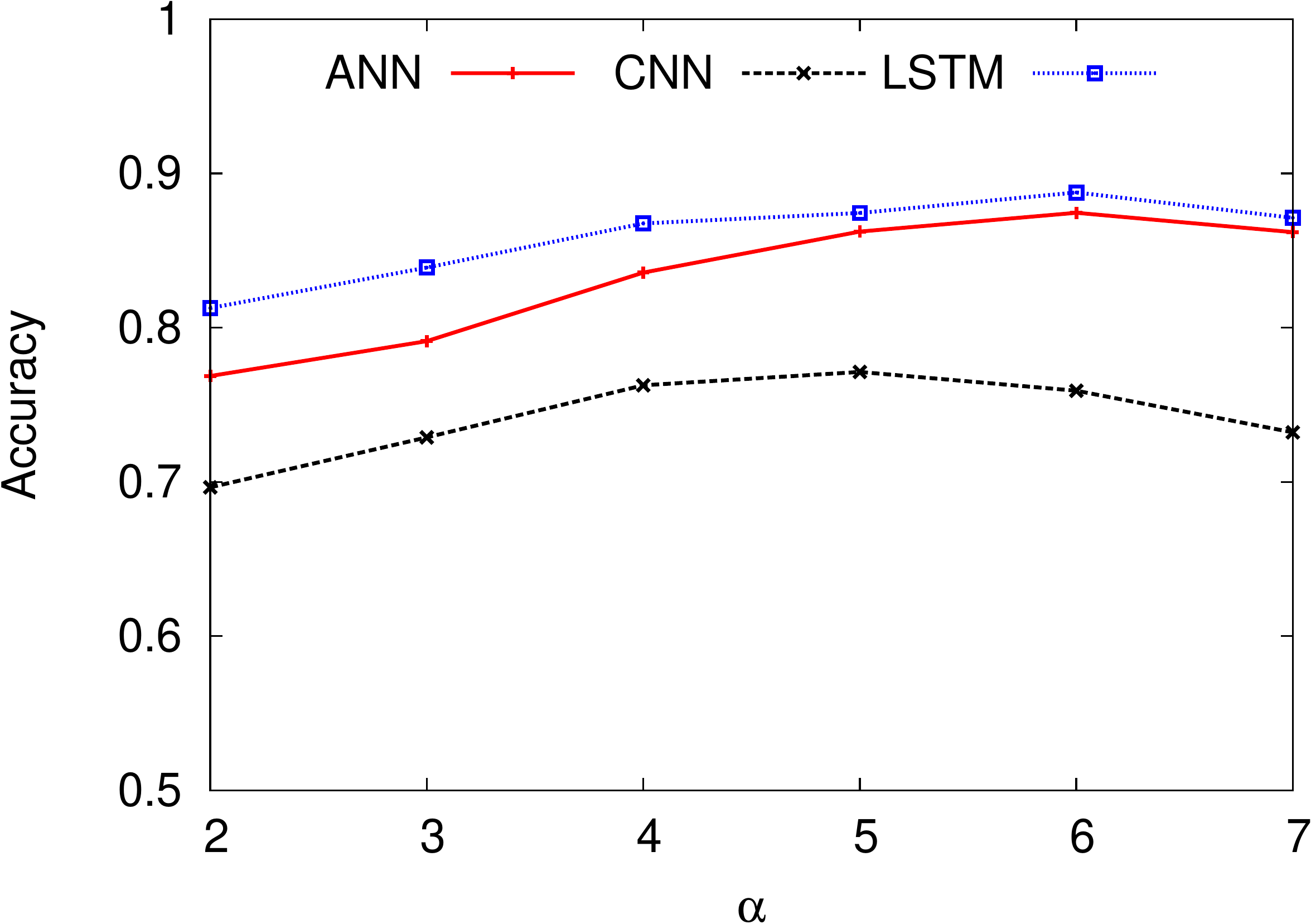}
  \caption{Impact of attacking traffic threshold.}
  \label{fig:alpha_impact}
\vspace{-1.5ex}
\end{figure}

As discussed previously, we need to set parameter $\alpha$ that indicates the number of times in the detection window the traffic appears as attacking traffic to decide whether the network is under attack or not. In this experiment, we evaluate the impact of $\alpha$ on the accuracy of the proposed approach. In Fig.~\ref{fig:alpha_impact}, we present the accuracy of the deep learning techniques for different values of $\alpha$. The results show that when $\alpha$ is too small or too large, the accuracy of the proposed approach degrades. Indeed, when $\alpha$ is too small, the detector will be very sensitive with the correlation among flows, leading to high number of false positives, i.e., normal traffic detected as attacking traffic. On the other hand, when $\alpha$ is too large, the detector is not sufficiently sensitive, leading to high number of false negatives. The results show that ANN and LSTM achieve the highest accuracy when $\alpha$ takes the value of $6$ while CNN has the best performance when $\alpha$ is equal to $5$. 

\begin{table}[t]
  \centering
  \caption{Training time and detection time (in seconds)}
\begin{tabular}{ |l| r| c | } 
\hline
 {\bf Techniques}  & {\bf Training time} & {\bf Detection time} \\
\hline
 ANN  & $58.69$  & $1.02$  \\
\hline
 CNN  & $92.24$  & $1.15$  \\
\hline
 LSTM  & $109.35$  & $1.10$ \\
\hline  
\end{tabular}
\label{tab:training_detection_time}  
\vspace{-2.5ex}  
\end{table}

\subsubsection{Training Time and Detection Time}
We also evaluate the training time of the learning model as well as its detection time for a set of data points. As shown in Table~\ref{tab:training_detection_time}, the training time of the proposed approach is only about $100$ seconds. LSTM spends the longest time for training the learning model. This could explain why it has the best performance. It is to be noted that this is a one time cost, given that historical traffic measurements are available to create a training dataset. In an online system, the training dataset could be enriched after new data points have been classified. The learning model can be updated with these new data points. The re-training process can be performed in parallel with the detection process. Thus, the training time will not cause any delay in detection time, allowing our approach to react against attacks promptly. It is also worth mentioning that detection time is also very short which is only $1$ second for all the deep learning techniques based on the proposed approach.

\section{Conclusion}  
\label{sec:conclusion}  
 
In this paper, we investigated the problem of coordinated/crossfire attacks in software-defined ITS networks. We proposed a machine learning based approach that leverages on the software-defined networking capability to capture traffic measurements such as the number of flows, aggregate flow size and timestamps of the traffic. We developed deep learning techniques using ANN, CNN and LSTM to train the learning model that can learn the temporal and spatial correlations among flows that originate from different compromised nodes. We evaluated the performance of the proposed approach by using Mininet-WiFi emulation platform. The experimental results show that the proposed approach achieves high performance in terms of accuracy, precision, recall and $F_1$-Score. The results also show that all the deep learning techniques used for training the proposed learning model achieve a detection accuracy of at least $80\%$. Among them, LSTM achieves the best performance with at least $87\%$ of detection accuracy. 

\section*{Acknowledgment}
  
This work was supported by Singapore MoE AcRF Tier-1 Grant, NUS WBS R-263-000-C13-112.

\bibliographystyle{IEEEtran}
\bibliography{vtc2019}	

\end{document}